\documentclass[aps,prd,twocolumn,groupedaddress,showpacs,showkeys]{revtex4-1}

\usepackage{atlasphysics}
\usepackage{graphicx}
\usepackage{amssymb}
\usepackage{array}
\usepackage{xspace}
\usepackage{enumerate}
\usepackage[hyperindex,breaklinks,colorlinks]{hyperref}

\def\chinoone{\ensuremath{\mathchoice%
      {\displaystyle\raise.4ex\hbox{$\displaystyle\tilde\chi_1$}}%
         {\textstyle\raise.4ex\hbox{$\textstyle\tilde\chi_1$}}%
       {\scriptstyle\raise.3ex\hbox{$\scriptstyle\tilde\chi_1$}}%
 {\scriptscriptstyle\raise.3ex\hbox{$\scriptscriptstyle\tilde\chi_1$}}}}

\def\mth{m_{3/2}}
\def\mdiff{\Delta m_{\chinoone}}
\def\nTRT{N_{\mathrm{TRT}}}

\def\trigdphi{\Delta \phi_{\mathrm{min}}^{\mathrm{jet\mathchar`-}\met}}

\def\nBlayer{N_{\mathrm{b\mathchar`-layer}}}
\def\nPixel{N_{\mathrm{pixel}}}
\def\nSCT{N_{\mathrm{SCT}}}
\def\probdis{\mathcal{P}^{\mathrm{dis}}_{\ell}}

\newcommand{\PaperTitle}{Search for charginos nearly mass-degenerate with the lightest neutralino based on a disappearing-track signature in $pp$ collisions at $\sqrt{s}=8~\mathrm{TeV}$ with the ATLAS detector}
\newcommand{\AbstractText}{ A search is presented for direct chargino production based on a disappearing-track signature using 20.3 $\ifb$ of proton--proton collisions at $\rts = 8~\TeV$ collected with the ATLAS experiment at the LHC.
In anomaly-mediated supersymmetry breaking (AMSB) models, the lightest chargino is nearly mass-degenerate with the lightest neutralino and its lifetime is long enough to be detected in the tracking detectors by identifying decays that result in tracks with no associated hits in the outer region of the tracking system.
Some models with supersymmetry also predict charginos with a significant lifetime.
This analysis attains sensitivity for charginos with a lifetime between $0.1~\mathrm{ns}$ and $10~\mathrm{ns}$, and significantly surpasses the reach of the LEP experiments.
No significant excess above the background expectation is observed for candidate tracks with large transverse momentum, and constraints on chargino properties are obtained.
In the AMSB scenarios, a chargino mass below $270~\GeV$ is excluded at 95\% confidence level.}

\usepackage{preprintcover}
\PreprintCoverPaperTitle{\PaperTitle}
\PreprintIdNumber{CERN-PH-EP-2013-155}
\PreprintCoverAbstract{\AbstractText}
\PreprintJournalName{Physical Review D}

\makeatletter
\g@addto@macro\bfseries{\boldmath}
\makeatother

\begin{document}

\title{\boldmath \PaperTitle}

\author{The ATLAS collaboration}
\email[]{atlas.publication@cern.ch}
\affiliation{CERN, 1211 Geneva 23, Switzerland}

\date{\today}

\begin{abstract}
  \AbstractText
\end{abstract}

%\keywords{}

\maketitle

\section{Introduction}\label{sec:introduction}
Anomaly-mediated supersymmetry breaking (AMSB) models~\cite{Giudice:1998xp,Randall:1998uk}, where soft supersymmetry (SUSY) breaking is caused by loop effects, provide a constrained mass spectrum of SUSY particles.
One prominent feature of these models is that the lightest supersymmetric particle is the nearly pure neutral wino that is mass-degenerate with the charged wino.
The lightest chargino ($\chinoonepm$) is then slightly heavier than the lightest neutralino ($\ninoone$) due to radiative corrections involving electroweak gauge bosons.
The typical mass splitting between $\chinoonepm$ and $\ninoone$ ($\mdiff$) is $\sim 160 \MeV$, which implies that $\chinoonepm$ has a considerable lifetime and predominantly decays into $\ninoone$ plus a low-momentum ($\sim 100 \MeV$) $\pi^{\pm}$.
The mean lifetime ($\tau_{\chinoonepm}$) of $\chinoonepm$ is expressed in terms of $\mdiff$ and expected to be typically a fraction of a nanosecond.
Several other SUSY models, which are motivated by the large value of the Higgs boson mass, also predict charginos with a significant lifetime and their decay to a soft pion and the lightest supersymmetric particle~\cite{Ibe:2012hu,Hall:2012zp,Arvanitaki:2012ps,ArkaniHamed:2012gw}.
Therefore, some charginos could have decay lengths exceeding a few tens of centimeters at the Large Hadron Collider (LHC).
When decaying in the sensitive volume, they are expected to be observed as ``disappearing tracks'' that have no more than a few associated hits in the outer region of the tracking system, and the softly emitted $\pi^{\pm}$ is not reconstructed.
This article explores AMSB scenarios by searching for charginos with their subsequent decays that result in such disappearing tracks.
The electroweak production of charginos has a sizable cross-section in proton--proton ($pp$) collisions at LHC energies.
Chargino-pair and chargino--neutralino associated production processes are identified using jets of large transverse momentum ($\pt$) from initial-state radiation ($pp \rightarrow \chinoonepm \ninoone j$ and $\chinoonep \chinoonem j$, where $j$ denotes a jet used to trigger the signal event).
The search presented here, based on $20.3~\ifb$ of $8 \TeV$ $pp$ collision data, increases the sensitivity compared to the previous ATLAS searches~\cite{ATLAS:2012ab,ATLAS:2012jp} due to analysis improvements and increases in the beam energy and luminosity.
The most significant improvement is achieved by enhancing the track reconstruction efficiency for charginos having short decay lengths.
In particular, the efficiency for charginos with $\tau_{\chinoonepm} \sim 0.2~\mathrm{ns}$, predicted for $\mdiff \sim 160~\mathrm{MeV}$, is around 100 times larger than in the previous searches.
The present analysis also provides sensitivity to a wider range of chargino lifetimes and covers a larger angular acceptance.
It significantly surpasses the reach of the LEP experiments~\cite{Heister:2002mn,Abdallah:2003gv,Acciarri:2000wy,Abbiendi:2002vz} for charginos with lifetimes $> 0.1~\mathrm{ns}$.

\section{The ATLAS detector}
ATLAS is a multi-purpose detector~\cite{Aad:2008zzm}, covering nearly the entire solid angle~\cite{Note1} around the collision point with layers of tracking devices surrounded by a superconducting solenoid providing a \mbox{2~T} axial magnetic field, a calorimeter system, and a muon spectrometer.
The inner detector (ID) provides track reconstruction in the region $|\eta| < 2.5$ and consists of pixel and silicon microstrip (SCT) detectors inside a straw-tube transition radiation tracker (TRT).
The pixel detector consists of three barrel layers and four disks in the forward and backward directions, providing on average three measurement points for charged tracks.
The SCT is composed of four cylindrical layers of double-sided silicon microstrip modules, with nine disk layers in each endcap region; eight silicon microstrip sensors are typically crossed by each track.
The TRT, of particular importance to this search, covers $|\eta|<1.0$ with its barrel detector, $0.8 < |\eta| < 2.0$ with the endcaps, and the radial range 563--1066 mm.
The average number of TRT hits on a track going through the inner detector in the central region is about 32.
Tracks in the transition region $0.8 < |\eta| < 1.2$ pass partially through both the barrel and endcap and are still expected to have $> 25$ hits on average.
Tracks passing through the dead region of the barrel TRT at $|\eta| < 0.1$ produce no TRT hits.
The calorimeter system covers the range of $|\eta| < 4.9$.
The electromagnetic calorimeter is a lead/liquid-argon (lead/LAr) detector in the barrel ($|\eta| < 1.475$) and endcap ($1.375 < |\eta| < 3.2$) regions.
The hadronic calorimeters are composed of a steel and scintillator barrel ($|\eta| < 1.7$), a copper/LAr endcap ($1.5 < |\eta| <3.2$), and a LAr forward system ($3.1< |\eta| <4.9$) with copper and tungsten absorbers.
The muon spectrometer consists of three large superconducting toroids, trigger chambers and precision tracking chambers that provide muon momentum measurements up to $|\eta|=2.7$.

\section{Data and simulated event samples}\label{sec:samples}
The data analyzed for this search were recorded in 2012 with the LHC colliding protons at $\rts = 8 \TeV$.
The integrated luminosity, after the application of beam, detector and data quality requirements, corresponds to $20.3 \pm 0.6 ~\ifb$, where the luminosity measurement is based on the calibration procedure described in Ref.~\cite{Aad:2013ucp} and uses the most recent van der Meer scans performed in November 2012 to determine the calibration and its uncertainty.

The analysis makes use of a dedicated topological trigger in order to suppress a huge Standard Model (SM) multijet background: it requires at least one jet with $\pt > 80~\GeV$, large missing transverse momentum (its magnitude, $\met$, above $70~\GeV$), and $\trigdphi > 1$, where $\trigdphi$ indicates the azimuthal separation between the missing transverse momentum and the jet.
If the event contains multiple jets with $\pt > 45 \GeV$, the smallest $\trigdphi$ value is taken by using either of the two highest-$\pt$ jets.
For the  multijet background, $\trigdphi$ peaks near zero since a large $\met$ is usually due to jet mismeasurement and is thus aligned with a high-$\pt$ jet, while the signal events cluster at $\trigdphi \approx \pi$.

Simulated Monte Carlo (MC) events are used to assess the experimental sensitivity to given models.
The minimal AMSB model is characterized by four parameters: the gravitino mass ($m_{3/2}$), the universal scalar mass ($m_0$), the ratio of Higgs vacuum expectation values at the electroweak scale ($\tan\beta$), and the sign of the higgsino mass term ($\mu$).
A large value of $1 \TeV$ is used for $m_0$ in order to prevent the appearance of a tachyonic slepton.
The production cross-section is determined largely by the wino mass and is fairly independent of the other parameters.
In this model, the wino mass is proportional to $m_{3/2}$.
The SUSY mass spectrum and the decay tables are calculated with the {\sc Isasusy} from {\sc Isajet} v7.80~\cite{Paige:2003mg}.
The corresponding MC signal samples are produced using Herwig++ 2.5.2~\cite{Bahr:2008pv} with {\sc CTEQ6L1}~\cite{Pumplin:2002vw} parton distribution functions of the proton (PDFs).
All samples used in this article are produced using a detector simulation~\cite{Aad:2010wqa} based on {\sc Geant4}~\cite{Agostinelli:2002hh}, and include multiple $pp$ interactions (pile-up) in the triggered and adjacent bunch crossings to model the pile-up effect.
Simulated points with chargino masses ($m_{\chinoonepm}$) ranging from $80\text{--}600 \GeV$ and various values of the chargino lifetime $\tau_{\chinoonepm}$ are generated.
In the {\sc Geant4} simulation the charginos decay exponentially and the branching fraction for the decay $\chinoonepm \to \ninoone \pi^{\pm}$ is set to 100\%.
Signal cross-sections are calculated at next-to-leading order in $\alpha_\mathrm{s}$ using {\sc Prospino2}~\cite{Beenakker:1996ch} program as shown in Fig.~\ref{figure:xsec_EW}.
The nominal cross-section and its uncertainty are taken from an envelope of cross-section predictions using different PDF sets and factorization and renormalization scales, as described in Ref.~\cite{Kramer:2012bx}.

\begin{figure}[ht!!]
  \centering
  \includegraphics[width=1.0\linewidth]{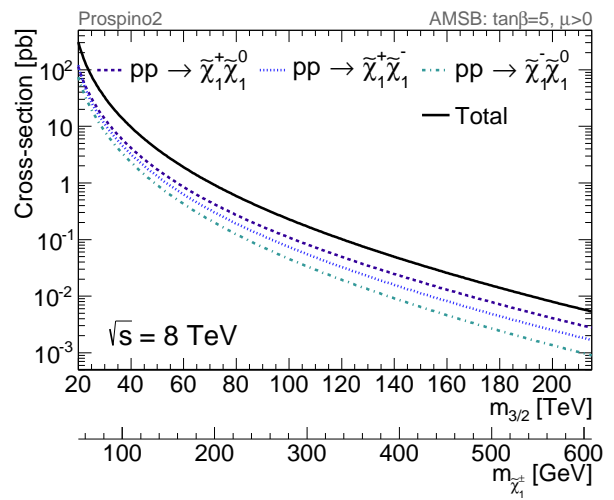}
  \caption{The cross-section for direct chargino production at $\rts = 8 \TeV$ as a function of the gravitino mass $\mth$.
  The corresponding chargino mass $m_{\chinoonepm}$ for each $\mth$ value is indicated.}
  \label{figure:xsec_EW}
\end{figure}

\section{Reconstruction, object identification and event selection}\label{sec:selection}
Standard Model processes, especially $W$+jet events that naturally have large $\met$, can result in final-state kinematics similar to that of the signal.
Kinematic selection criteria are applied to ensure high trigger efficiency and to reduce background arising from multijet processes or from electroweak gauge bosons that decay leptonically.
At the next stage, the vast majority of SM background events are removed by identifying and demanding a disappearing track in the event.

\subsection{Track reconstruction} \label{sec:tracking}

Charged particle trajectories are reconstructed as tracks in the ID.
In order to improve the efficiency of reconstructing particles leaving short tracks, this analysis applies an extra extended-track reconstruction that provides pixel-seeded reconstructed tracks in addition to the ATLAS standard tracks.
The standard track reconstruction algorithm~\cite{Cornelisen:2008} is a sequence made of two main steps.
First, an inside-out sequence starts from triplets of three-dimensional space points from the pixel and SCT detectors, with each space point originating from a unique detector layer, and then extends the resulting trajectories by combining other pixel, SCT and TRT hits.
A second sequence takes the remaining TRT hits as seeds and attempts to extend identified trajectories inwards by combining them with unused space points.
The first sequence is optimized to find primary tracks coming from the interaction point, while the second sequence is optimized for the reconstruction of electrons from photon conversions in the ID volume.
The inside-out sequence is of particular interest for finding long-lived chargino trajectories, although it is optimized for the reconstruction of stable particles that leave long tracks in the ID, and in particular, only reconstructs tracks with a minimum of seven space points.
In order to increase the acceptance of the track reconstruction and especially the chargino track reconstruction efficiency at low radius, a third sequence is applied. This sequence proceeds using leftover pixel and SCT hits from the two previous tracking sequences and reconstructs tracks with a minimum of three pixel hits, while no SCT or TRT hits are required.
The outward extension then follows; SCT and TRT hits are attached if they lie along the track trajectory.
The tracks reconstructed by the third sequence are used only to select disappearing-track candidates.

\subsection{Event reconstruction} \label{sec:object}
The event vertex~\cite{ATLAS-CONF-2010-069} is required to have at least five associated tracks. When more than one such vertex is found, the vertex with the largest $\sum |\pt|^2$ of the associated tracks is chosen as primary.
Jets are reconstructed using the anti-$k_{t}$ algorithm~\cite{Cacciari:2008gp} with a distance parameter of 0.4.
The inputs to the jet reconstruction algorithm are topological calorimeter energy clusters seeded by cells with energy significantly above the noise level.
Jet energies are then calibrated back to the particle level~\cite{Aad:2011he}.
Reconstructed jets must satisfy the requirements of $\pt > 20 \GeV$ and $|\eta| < 2.8$.
Electron candidates are reconstructed from energy clusters in the electromagnetic calorimeter matched to a track in the ID.
Electrons must then fulfill the ``loose'' identification requirements described in Ref.~\cite{Aad:2011mk}, have transverse energy $\et > 10 \GeV$, and be within the region $|\eta| < 2.47$.
Muon candidates are formed by matching ID tracks with either a complete track or a track segment reconstructed in the muon spectrometer~\cite{ATLAS-CONF-2011-063}.
Furthermore, muons must satisfy the requirements of $\nBlayer > 0$ if crossing an active module of the innermost pixel layer, $\nPixel > 0$, $\nSCT \ge 6$, $\pt > 10 \GeV$, and $|\eta| < 2.4$, where $\nBlayer$, $\nPixel$ and $\nSCT$ are the numbers of hits in the innermost pixel layer, the pixel and SCT detectors, respectively.

Following the object reconstruction described above, overlaps between jets and leptons are resolved to ensure isolation of leptons.
First, any jet candidate lying within a distance of $\Delta R \equiv \sqrt{\left( \Delta\eta \right)^2 + \left( \Delta\phi \right)^2} = 0.2$ of an electron is discarded.
Then, any lepton candidate within a distance of $\Delta R = 0.4$ of any surviving jet is discarded.

The calculation of $\met$ is based on the transverse momenta of remaining jets and lepton candidates and on all calorimeter energy clusters that are not associated with such objects~\cite{Aad:2012re}.

\subsection{Kinematic selection} \label{sec:kinematic_selection}
Following the event reconstruction, selection requirements to reject non-collision background events, given in Ref.~\cite{Aad:2011he}, are applied to jets.
In order to suppress backgrounds from $W/Z+\text{jets}$ and top-pair production processes, events are discarded if they contain any electron or muon candidates (lepton veto).
Events containing muons are further suppressed by requiring no tracks with $\pt > 10~\GeV$ reconstructed in the muon spectrometer.
The candidate events are finally required to have $\met > 90 \GeV$, at least one jet with $\pt > 90 \GeV$, and $\trigdphi > 1.5$.
The trigger selection is $>98\%$ efficient for signal events satisfying these selection requirements.

\subsection{Selection of disappearing tracks} \label{sec:track_selection}
The tracks originating from charginos are expected to have high transverse momenta, to be isolated, and to have few associated hits in the outer region of the ID.
The TRT detector in particular provides substantial discrimination against penetrating stable charged particles if only a small number of hits on the track is required.
Therefore, candidate tracks for decaying charginos are required to fulfill the following criteria:
\begin{enumerate}[{(}I{)}]
  \item the track must have $\nPixel \ge 3$, $\nBlayer \ge 1$ if crossing an active module of the innermost pixel layer, $\nSCT \ge 2$, $|d_0| < 0.1~\mathrm{mm}$ and $|z_0\sin\theta| < 0.5~\mathrm{mm}$, where $d_0$ and $z_0$ are the transverse and longitudinal impact parameters with respect to the primary vertex;
  \item the track reconstruction must be of good quality, meeting the following requirements: it must have a track fit $\chi^2$-probability of $>10\%$, no hits formed in a single pixel row of which the readout is shared with another pixel, and no hits missing in active silicon modules along the trajectory between the first and last hit of the track;
  \item the track must be isolated: it must fulfill $\pt^{\mathrm{cone40}}/\pt < 0.04$, where $\pt^{\mathrm{cone40}}$ is the sum of $\pt$ of all tracks with $\pt > 400~\MeV$, $|d_0|<1.5~\mathrm{mm}$, and $|z_0\sin\theta|<1.5~\mathrm{mm}$ that lie within a cone of $\Delta R = 0.4$ around the track. There must also be no jets having $\pt$ above $45 \GeV$ within a cone of $\Delta R = 0.4$ around the candidate track;
  \item the candidate track must have $\pt$ above $15 \GeV$, and must be the highest-$\pt$ isolated track in the event;
  \item the candidate track must satisfy $0.1 < |\eta| < 1.9$; 
  \item the number of TRT hits associated with the track ($\nTRT$), determineed by counting hits lying on the extrapolated track, must be less than five.
\end{enumerate}
Criteria (I) and (II) are applied in order to ensure well-reconstructed primary tracks.
Criteria (III) and (IV) are employed to select chargino tracks that are isolated and have in most cases the highest $\pt$.
These criteria also substantially reduce background tracks from the pile-up.
Criterion (V) is used to ensure coverage by the TRT active region and enhance the rejection of background tracks.
Criterion (VI) helps to remove the majority of background tracks in SM processes, as shown in Fig.~\ref{figure:nTRT}. For SM charged particles traversing the TRT detector, the number of TRT hits is typically $\nTRT \simeq 32$, whereas for charginos that decay before reaching the TRT the expected is $\nTRT \simeq 0$.
Hereafter, ``high-$\pt$ isolated track selection'' and ``disappearing-track selection'' indicate criteria (I)--(V) and (I)--(VI), respectively.

Making use of short tracks and the whole TRT detector for the background track rejection extends the sensitive decay volume inwards and enlarges the region of signal acceptance in $\eta$.
This results in better sensitivity for charginos, especially with small lifetime, than in the previous search~\cite{ATLAS:2012jp} based on the $7 \TeV$ collision data.
Figure~\ref{figure:C1TrkIdEff} shows the tracking efficiency with the disappearing-track selection for decaying charginos as a function of the radius and $\eta$ of the decay vertex.
It is fully efficient for charginos that reach the first SCT layer and decay before reaching the TRT detector. The MC simulation shows that it is also largely independent of $m_{\chinoonepm}$.
A summary of the kinematic selection criteria, disappearing-track requirements, and data reduction is given in Table~\ref{table:data_reduction}.

\begin{figure}[ht!!]
  \centering
  \includegraphics[width=1.0\linewidth]{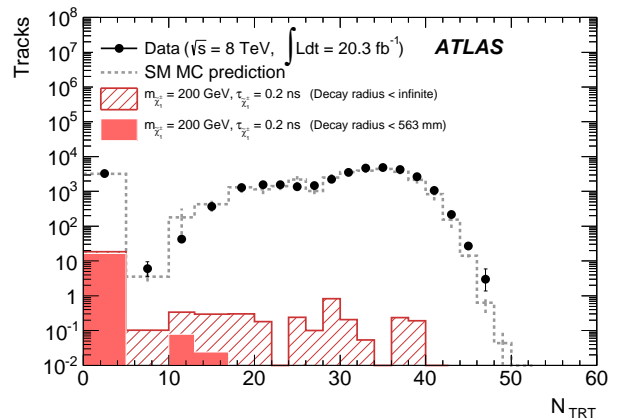}
  \caption{Number of TRT hits ($\nTRT$) for data and signal MC events ($m_{\chinoonepm}$ = $200 \GeV$, $\tau_{\chinoonepm}$ = $0.2~\mathrm{ns}$) with the high-$\pt$ isolated track selection.
  The expectation from SM MC events is also shown.
  The solid colored histogram shows the expected distribution for charginos with a decay radius $< 563~\mathrm{mm}$ while the hatched histogram shows it for all charginos for these mass and lifetime values. Tracks with $\nTRT < 5$ in SM events, mimicking the decaying-chargino signature, are described in Sec.~\ref{sec:background}.}
    \label{figure:nTRT}
\end{figure}

\begin{figure}[ht!!]
  \centering
  \includegraphics[width=1.0\linewidth]{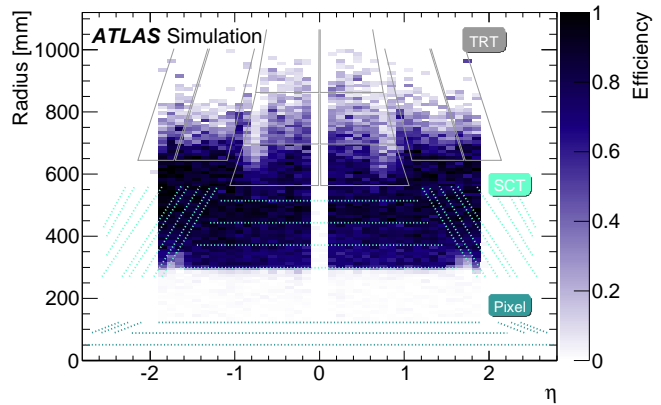}
  \caption{The efficiency for decaying charginos with the disappearing-track selection.
  Vertical and horizontal axes are the radius and $\eta$ of the decay, respectively.
  Sensitive layers and areas of the pixel, SCT and TRT detectors are also indicated in the figure.
  }
    \label{figure:C1TrkIdEff}
\end{figure}

\begin{table*}[ht!!]
  \centering
  \caption{Summary of selection requirements and data reduction for data and expected signal events ($m_{\chinoonepm}$ = $200 \GeV$, $\tau_{\chinoonepm}$ = $0.2~\mathrm{ns}$). The signal selection efficiencies are also shown in parentheses. Signal efficiencies are low at the first stage due to the trigger based on a jet from initial-state radiation.}
  \begin{tabular*}{\textwidth}{@{\extracolsep{\fill}}crr@{}}
    \hline
    \hline
    Selection requirement & Observed events & Expected signal MC events (efficiency~[\%]) \\
    \hline
    Quality requirements and trigger           & 20479553 & 1873 (8.8)  \\
    Jet cleaning                               & 18627508 & 1867 (8.8)  \\
    Lepton veto                                & 12485944 & 1827 (8.6)  \\
    Leading jet $\pt > 90 \GeV$                & 10308840 & 1571 (7.4)  \\
    $\met > 90 \GeV$                           & 6113773 & 1484 (7.0)  \\
    $\trigdphi > 1.5$                          & 5604087 & 1444 (6.8)  \\
    High-$\pt$ isolated track selection        & 34379 & 21.9 (0.10)  \\
    Disappearing-track selection               & 3256 & 18.4 (0.087)  \\
    \hline
    \hline
  \end{tabular*}
  \label{table:data_reduction}
\end{table*}

\section{Estimate of the $\pt$ spectrum of background tracks} \label{sec:background}
There are three primary sources of tracks from background processes that mimic the disappearing-track signature: charged hadrons interacting with material in the ID (interacting-hadron tracks), prompt electrons or muons failing to satisfy their identification criteria (lepton tracks) and low-$\pt$ charged particles whose $\pt$ is highly mismeasured ($\pt$-mismeasured tracks).
Interacting-hadron and electron tracks are responsible for the background in the approximate range $\pt < 50~\GeV$, whereas $\pt$-mismeasured tracks are dominant for $\pt > 100~\GeV$.
A small contribution from muon tracks is expected throughout the full $\pt$ range.
The contribution of charged-hadron decays is significantly smaller than that of interacting hadrons; therefore, such a background source is neglected.
A background estimation based on the MC simulation has difficulty in accurately describing the properties of these background tracks.
Therefore, the background contribution to the disappearing-track candidates is estimated using techniques that do not rely on the MC simulation.
Each of the three types of background tracks shows a distinctive $\pt$ spectrum; a simultaneous fit is performed for signal and background yields using the observed $\pt$ spectrum and templates of background-track $\pt$ spectra produced from dedicated control data samples.
The $\pt$ spectra of the first two background types are obtained in the same way as in Ref.~\cite{ATLAS:2012jp}.

\subsection{Interacting-hadron tracks}
Charged hadrons, mostly charged pions, can interact with material in the ID and their tracks can be misidentified as disappearing tracks.
The shape of the $\pt$ distribution of interacting-hadron tracks is obtained from that of non-interacting-hadron tracks.
In the $\pt$ range above $15 \GeV$, where inelastic interactions dominate, the interaction rate has nearly no dependence on $\pt$~\cite{Nakamura:2010zzi}, which is also confirmed by the detector simulation.
By adopting kinematic selection criteria identical to those for the signal and ensuring traversal of the TRT detector by requiring $\nTRT>25$, a data sample of non-interacting-hadron tracks is obtained.
A pure control data sample is ensured by requiring associated calorimeter activity and removing the contamination from electron and muon tracks (described below) and any chargino signal.
The following requirements are applied: $\ET^{\mathrm{cone40}}$ $>$ $7.5~\GeV$ and $\sum_{\Delta R<0.4} \ET^{\mathrm{clus}}/\pt^{\mathrm{track}} > $ 0.4, where $\ET^{\mathrm{cone40}}$ is the calorimeter transverse energy deposited in a cone of $\Delta R<0.4$ around the track (excluding $\ET$ of the calorimeter cluster matched to the track), $\sum_{\Delta R<0.4} \ET^{\mathrm{clus}}$ is the sum of cluster energies in a cone of $\Delta R < 0.4$ around the track, and $\pt^{\mathrm{track}}$ is the track $\pt$.

In most cases, interacting hadrons have associated calorimeter activity that can be used to form jets.
Therefore, after the selection requirements, the contribution of this background to the disappearing-track candidates having $\pt > 100~\GeV$ is negligibly small.

\subsection{Leptons failing to satisfy identification criteria} \label{sec:electron}
Some charged leptons ($\ell \equiv e~\text{or}~\mu$) lose much of their momenta in the ID due to scattering with material or large bremsstrahlung.
Such leptons are unlikely to be correctly identified (hence surviving the lepton veto) and may be classified as disappearing tracks.

In order to estimate the lepton-track background, a control data sample is defined by requiring kinematic selection identical to those for the signal search sample, while requiring one lepton that fulfills both its identification criteria and the isolated track selection criteria.
The $\pt$ spectrum of leptons without any identification requirements is obtained by applying a correction for the identification efficiency.
The $\pt$ distribution of lepton background tracks is then estimated by multiplying this distribution by the probability ($\probdis$) of failing to satisfy the lepton identification criteria (hence being retained in the signal search sample) and passing the disappearing-track selection criteria.
The electron and muon components are considered separately.

For the measurement of $\probdis$, a tag-and-probe method is applied to $\Zll$ events collected with unprescaled single-lepton triggers and by requiring a $Z$ boson candidate with reconstructed invariant mass within $\pm 5~\GeV$ of the $Z$ mass.
Tag-leptons are required to be well isolated from jets and to fulfill the lepton identification criteria.
Probe-leptons are selected without any identification requirements but with exactly the same high-$\pt$ isolated track selection criteria used for chargino candidate tracks.
The probability $\probdis$ is given by the fraction of events in which the probe-lepton passes the disappearing-track selection criteria; it ranges between $10^{-2}$--$10^{-4}$ for electrons and $10^{-4}$--$10^{-5}$ for muons.
Statistical uncertainties and uncertainties on the identification efficiency are considered in deriving the estimated $\pt$ spectra and their uncertainties.

\subsection{Tracks with mismeasured $\pt$} \label{sec:fakes}
The background contribution to disappearing-track candidates with $\pt > 100~\GeV$ originates primarily from tracks with mismeasured $\pt$ ($\pt$-mismeasured tracks).
A high density of silicon hits, hadronic interactions and scattering can lead to combinations of wrong space-points in the procedure of track-seed finding or outward-extension of trajectories, resulting in anomalously high values of $\pt$ especially for short-length tracks.
Simulation studies indicate that the $\pt$ spectrum of such tracks depends little on the reconstructed $d_0$ or production process.
Figure~\ref{figure:faketrack_pt_qcd} shows the $\pt$ spectrum of disappearing tracks with different $d_0$ values in a multijet-enriched data sample collected with single-jet triggers and requirements of $\met < 90~\GeV$ and no leptons: the contamination from interacting-hadron and lepton tracks is expected to be very small in the range $\pt > 50~\GeV$ or $|d_0| > 1~\mathrm{mm}$.
The $\pt$ shape of $\pt$-mismeasured tracks with $|d_0| < 0.1~\mathrm{mm}$ is found to be the same as that of similarly mismeasured tracks with $1~\mathrm{mm} < |d_0| < 10~\mathrm{mm}$.
A sample with a nearly pure $\pt$-mismeasured track contribution can be obtained with the same requirements as for the signal tracks, while requiring $1~\mathrm{mm} < |d_0| < 10~\mathrm{mm}$.
The $\pt$ shape is finally determined by a fit to the sample by a functional form $x^{-a}$ ($x \equiv \pt^{\mathrm{track}}$), where $a = 1.78 \pm 0.05$ is obtained.

\begin{figure}[ht!!]
  \centering
  \includegraphics[width=1.0\linewidth]{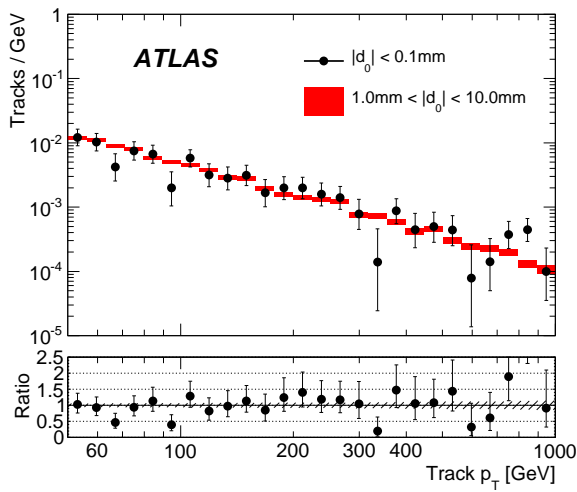}
  \caption{The $\pt$ distributions of disappearing tracks with impact parameter ranges $|d_0| < 0.1~\mathrm{mm}$ and $1~\mathrm{mm} < |d_0| < 10~\mathrm{mm}$ in the multijet-enriched data sample, normalized to unity.
  The ratio between the two distributions is also shown at the bottom of the figure.
  The error bars and band in the ratio plot indicate the statistical uncertainties of each sample.
  }
  \label{figure:faketrack_pt_qcd}
\end{figure}

\section{Estimate of systematic uncertainties}\label{sec:uncertainties}
The sources of systematic uncertainty on the signal expectation are the following:
the theoretical cross-section, parton radiation model, jet energy scale (JES) and resolution (JER), trigger efficiency, pile-up modeling, track reconstruction efficiency, and the integrated luminosity.
The contributions of each systematic uncertainty in the signal yield are summarized in Table~\ref{table:summary_uncertainty} for two reference signal samples.

Theoretical uncertainties on the signal cross-section, already described in Sec.~\ref{sec:samples}, range from 6\% to 8\% depending on $m_{\chinoonepm}$.
The uncertainties on the modeling of high-$\pt$ jets, originating from initial- and final-state radiation, are estimated by varying the generator tunes in the simulation as well as by generator-level studies carried out on samples produced with an additional jet in the matrix-element method using {\sc MadGraph5} program~\cite{Alwall:2011uj} and {\sc Pythia6} program~\cite{Sjostrand:2006za}.
By adopting PDF tunes that provide less and more radiation and taking the maximum deviation from the nominal tune, the uncertainty due to jet radiation is evaluated.
The uncertainty arising from the matching of matrix elements with parton showers is evaluated by doubling and halving the default value of the matching parameter~\cite{Alwall:2007fs}.
The resulting changes are combined in quadrature and yield an uncertainty of 10--17\% depending on $m_{\chinoonepm}$.
The uncertainties on the JES and JER result in a variation of the signal selection efficiency that is assessed according to Ref.~\cite{Aad:2011he}, and an uncertainty of 3--6\% is assigned.
An uncertainty due to the trigger efficiency is estimated to be 4.5\% by taking the difference between data and MC simulation in a $W$+jet sample in which $W$ decays into $\mu$ plus $\nu_{\mu}$.
The uncertainty originating from the pile-up modeling in the simulation is evaluated by weighting simulated samples so that the average number of pile-up interactions is varied by $\pm$ 10\%, which yields a $0.5\%$ uncertainty on the signal efficiency.
The ID material affects the track reconstruction efficiency.
An uncertainty of $2\%$ is assigned from Ref.~\cite{Aad:2011qe} to take into account differences in the tracking efficiency between data and MC simulation related to the detector material description in the simulation. 
The uncertainty on the integrated luminosity is $\pm 2.8\%$. It is derived, following the same methodology as that detailed in Ref.~\cite{Aad:2013ucp}. %, from a preliminary calibration of the luminosity scale obtained from beam-separation scans performed in November 2012.

Systematic uncertainties on the background $\pt$ shapes and normalizations arising from statistical uncertainties of the control data samples and uncertainties on the lepton identification efficiencies are also considered in deriving the results (discussed in Sec.~\ref{sec:fit}).
In order to account for a possible bias induced by the $d_0$ requirement in the control data sample of $\pt$-mismeasured tracks, an additional uncertainty is assigned by taking the difference between the value of the parameter $a$ given in Sec.~\ref{sec:fakes} and the value $1.82 \pm 0.07$ derived using SM background MC events remaining after the selection requirements.

\begin{table}[ht!!]
  \centering
  \caption{Summary of systematic uncertainties [\%] on the expected number of signal events for $m_{\chinoonepm} = 200~\GeV$ and $300~\GeV$.}
  \begin{tabular*}{\linewidth}{@{\extracolsep{\fill}}lrr@{}}
    \hline
    \hline
    & \mbox{200~\GeV} & \mbox{300~\GeV} \\
    \hline
    \multicolumn{3}{l}{(\bf{Theoretical uncertainty})} \\
    Cross-section       & 6.4 & 6.8 \\
    \hline
    \multicolumn{3}{l}{(\bf{Uncertainty on the acceptance})} \\
    Modeling of initial/final-state radiation & 14.5 & 16.4 \\
    JES/JER                                   & 3.9  & 6.0 \\
    Trigger efficiency                        & 4.5  & 4.5 \\
    Pile-up modeling                         & 0.5  & 0.5 \\
    Track reconstruction efficiency           & 2.0  & 2.0 \\
    Luminosity                                & 2.8  & 2.8 \\
    {\bf Sub-total} & 16.1 & 18.4 \\
    \hline
    \hline
  \end{tabular*}
  \label{table:summary_uncertainty}
\end{table}

\section{Fit to the $\pt$ spectrum of disappearing tracks} \label{sec:fit}

\begin{table*}[ht!!]
  \centering
  \caption{Numbers of observed and expected background events as well as the probability that a background-only experiment is more signal-like than observed ($p_0$) and the model-independent upper limit on the visible cross-section ($\sigma_{\mathrm{vis}}^{95\%}$) at 95\% CL.}
  \begin{tabular*}{\textwidth}{@{\extracolsep{\fill}}ccccc@{}}
    \hline
    \hline
     & $\pt^{\mathrm{track}}>75~\GeV$ & $\pt^{\mathrm{track}}>100~\GeV$ & $\pt^{\mathrm{track}}>150~\GeV$ & $\pt^{\mathrm{track}}>200~\GeV$ \\
    \hline
    Observed events                              & 59              & 36             & 19             & 13 \\
    Expected events                              & $48.5 \pm 12.3$ & $37.1 \pm 9.4$ & $24.6 \pm 6.3$ & $18.0 \pm 4.6$ \\
    $p_0$ value                                  & 0.17            & 0.41           & 0.46           & 0.44 \\
    Observed $\sigma_{\mathrm{vis}}^{95\%}~\mathrm{[fb]}$ & 1.76                     & 1.02                &  0.62                           & 0.44 \\
    Expected $\sigma_{\mathrm{vis}}^{95\%}~\mathrm{[fb]}$ & $1.42^{+0.50}_{-0.39}$  & $1.05^{+0.37}_{-0.28}$   &  $0.67^{+0.27}_{-0.19}$   & $0.56^{+0.23}_{-0.16}$ \\
    \hline
    \hline
  \end{tabular*}
  \label{table:pvalue}
\end{table*}

The signal hypothesis with a given value of $m_{\chinoonepm}$ and $\tau_{\chinoonepm}$ is tested based on an extended maximum likelihood fit to the $\pt$ spectrum of the disappearing-track candidates.
The likelihood function for the track $\pt$ consists of one probability density function for the signal and four for the different backgrounds derived in Sec.~\ref{sec:background}. 
In the fit, the yields of the signal, interacting-hadron, and $\pt$-mismeasured tracks are left free.
The yields of electron and muon background tracks are constrained to their estimated values within the uncertainties.
The effects of systematic uncertainties on the yields and the parameters describing the $\pt$-distribution shapes of the background tracks are also incorporated into the likelihood function.

The number of observed events having a high-$\pt$ disappearing track above a given threshold and the expectation for the background, derived by the background-only fit in the $\pt$ range below $75~\GeV$, are given in Table~\ref{table:pvalue}.
No significant deviations from the background expectations are found.
The probability ($p_0$ value) that a background-only experiment is more signal-like than the observation and the model-independent upper limit on the visible cross-section ($\sigma_\mathrm{vis}^{95\%}$) at 95\% confidence level (CL) are also given in the table.
Figure~\ref{figure:trackpt_bestfit} shows the $\pt$ distribution for the selected data events compared to the background model derived by the background-only fit in the full $\pt$ range: the best-fit values for the yields of interacting hadrons, electron tracks, muon tracks and $\pt$-mismeasured tracks are $2187 \pm 71$, $852 \pm 35$, $23 \pm 8$ and $212 \pm 33$, respectively.
Three selected examples for the signal are also shown in the figure.

An excess with a corresponding significance of $\sim 2\sigma$ is seen in Fig.~\ref{figure:trackpt_bestfit} at $\pt$ around $90~\GeV$.
Detailed investigation of the events in this region show no peculiarities or significant differences in event kinematics or track properties compared to candidates in nearby track-$\pt$ regions.
The discrepancy is also not consistent with any of the signal hypotheses studied in this article.
For the models considered, high-$\pt$ tracks are expected and the best expected sensitivity derives from the region with $\pt$ above $200~\GeV$, where a deficit is observed as reported in Table~\ref{table:pvalue}.

Events with two disappearing-track candidates, being particularly sensitive to chargino-pair production with a long lifetime, are also explored.
One candidate event is found; however, the event lacks high-$\pt$ disappearing-track candidates (their $\pt$ being $30~\GeV$ and $18~\GeV$).

\begin{figure}[ht!!]
  \centering
  \includegraphics[width=\linewidth]{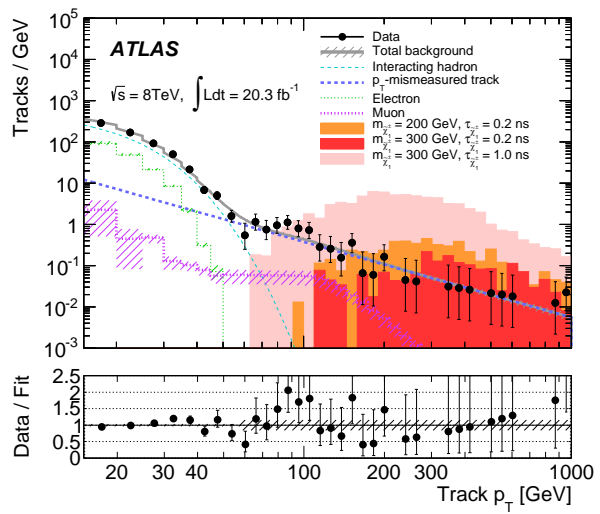}
  \caption{The $\pt$ distribution of disappearing-track candidates. The solid circles show data and lines show each background track-$\pt$ spectrum obtained by the background-only fit.
  The resulting uncertainties on the $\pt$ spectrum for each background are indicated by the error bands.
  The signal expectations are also shown.
  The ratio of the data to the background track-$\pt$ spectrum is shown at the bottom of the figure.
  }
  \label{figure:trackpt_bestfit}
\end{figure}

\section{Results}\label{sec:results}
In the absence of a signal, constraints are set on $m_{\chinoonepm}$ and $\tau_{\chinoonepm}$.
The upper limit on the production cross-section for a given $m_{\chinoonepm}$ and $\tau_{\chinoonepm}$ at 95\% CL is set at the point where the CL of the ``signal+background'' hypothesis, based on the profile likelihood ratio~\cite{Cowan:2010js} and the $CLs$ prescription~\cite{Read:2002hq}, falls below 5\% when scanning the CL along various values of signal strength.
The constraint on the allowed $\tau_{\chinoonepm}\text{--}m_{\chinoonepm}$ parameter space is shown in Fig.~\ref{figure:limit_lifetime_mC1}.
The expected limit is set by the median of the distribution of 95\% CL limits calculated by pseudo-experiments with the expected background and no signal, where the systematic parameters are varied according to their systematic uncertainties.
The regions excluded by the previous ATLAS search~\cite{ATLAS:2012jp} and the LEP2 searches are indicated.
The example of the exclusion reached by the ALEPH experiment~\cite{Heister:2002mn} of $88~\GeV$ at 95\% CL that is derived for the chargino mass in the case of heavy sfermions, irrespective of the chargino-neutralino mass difference is shown as the LEP2 result. This constraint is largely independent of $\tan\beta$ or the sign of $\mu$.

The analysis is not performed for signals having $\tau_{\chinoone} > 10~\mathrm{ns}$ (corresponding $\Delta m_{\chinoone}$ being below the charged pion mass) because a significant fraction of charginos would traverse the ID before decaying, thereby reducing the event selection efficiency.
In these scenarios the charginos are considered as stable particles and the main search tool would be to look for tracks with anomalous ionization energy loss~\cite{Aad:2012pra}.
In comparison with the previous result, the sensitivity to charginos having $\tau_{\chinoonepm} < 1~\mathrm{ns}$ is significantly improved and the exclusion reach is extended by $\sim 200~\GeV$.

Figure~\ref{figure:limit_dM_mC1} shows the constraint on the allowed $\mdiff\text{--}m_{\chinoonepm}$ parameter space of the minimal AMSB model; the expected 95\% CL exclusion reaches $m_{\chinoonepm}=245^{+25}_{-30}~\GeV$ for $\mdiff \sim 160~\mathrm{MeV}$.
The limits on $\tau_{\chinoonepm}$ are converted into limits on $\mdiff$ following Ref.~\cite{Chen:1999yf}.
The theoretical prediction of $\mdiff$ for wino-like lightest chargino and neutralino states at two-loop level~\cite{Ibe:2012sx} is also indicated in the figure.
A new limit that excludes charginos of $m_{\chinoonepm}< 270 \GeV$ (corresponding $\mdiff$ and $\tau_{\chinoonepm}$ being $\sim 160~\MeV$ and $\sim 0.2~\mathrm{ns}$, respectively) at 95\% CL is set in the AMSB models.

\begin{figure}[ht!!]
  \centering
  \includegraphics[width=1.0\linewidth]{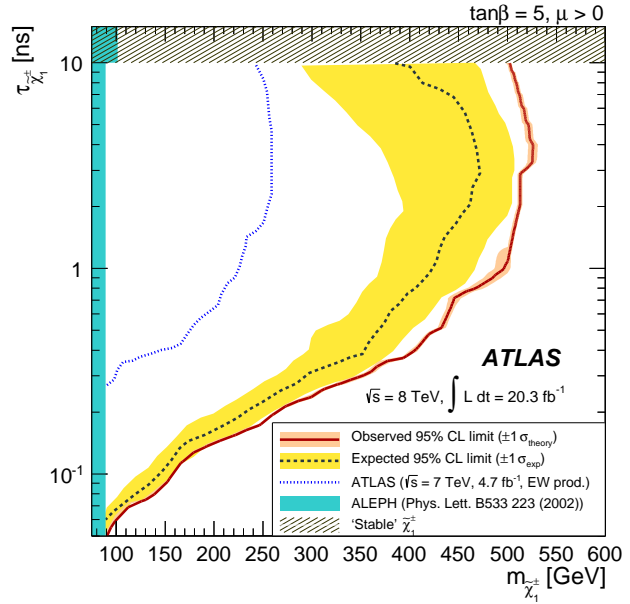}
  \caption{The constraint on the allowed $\tau_{\chinoonepm}\text{--}m_{\chinoonepm}$ space for $\tan\beta = 5$ and $\mu>0$.
  The black dashed line shows the expected limits at 95\% CL, with the surrounding shaded band indicating the 1$\sigma$ exclusions due to experimental uncertainties.
  Observed limits are indicated by the solid bold contour representing the nominal limit and the narrow surrounding shaded band is obtained by varying the cross-section by the theoretical scale and PDF uncertainties.
  The previous result from Ref.~\cite{ATLAS:2012jp} and an example of the limits achieved at LEP2 by the ALEPH experiment~\cite{Heister:2002mn} are also shown on the left by the dotted line and the shaded region, respectively.
  The search for charginos with long lifetimes, as indicated by the upper shaded region, is not covered by this analysis. The limits achieved at LEP2 by the ALEPH experiment of $101\GeV$ for long-lived charginos is taken from ~\cite{Heister:2002mn}.
  }
  \label{figure:limit_lifetime_mC1}
\end{figure}

\begin{figure}[ht!!]
  \centering
  \includegraphics[width=1.0\linewidth]{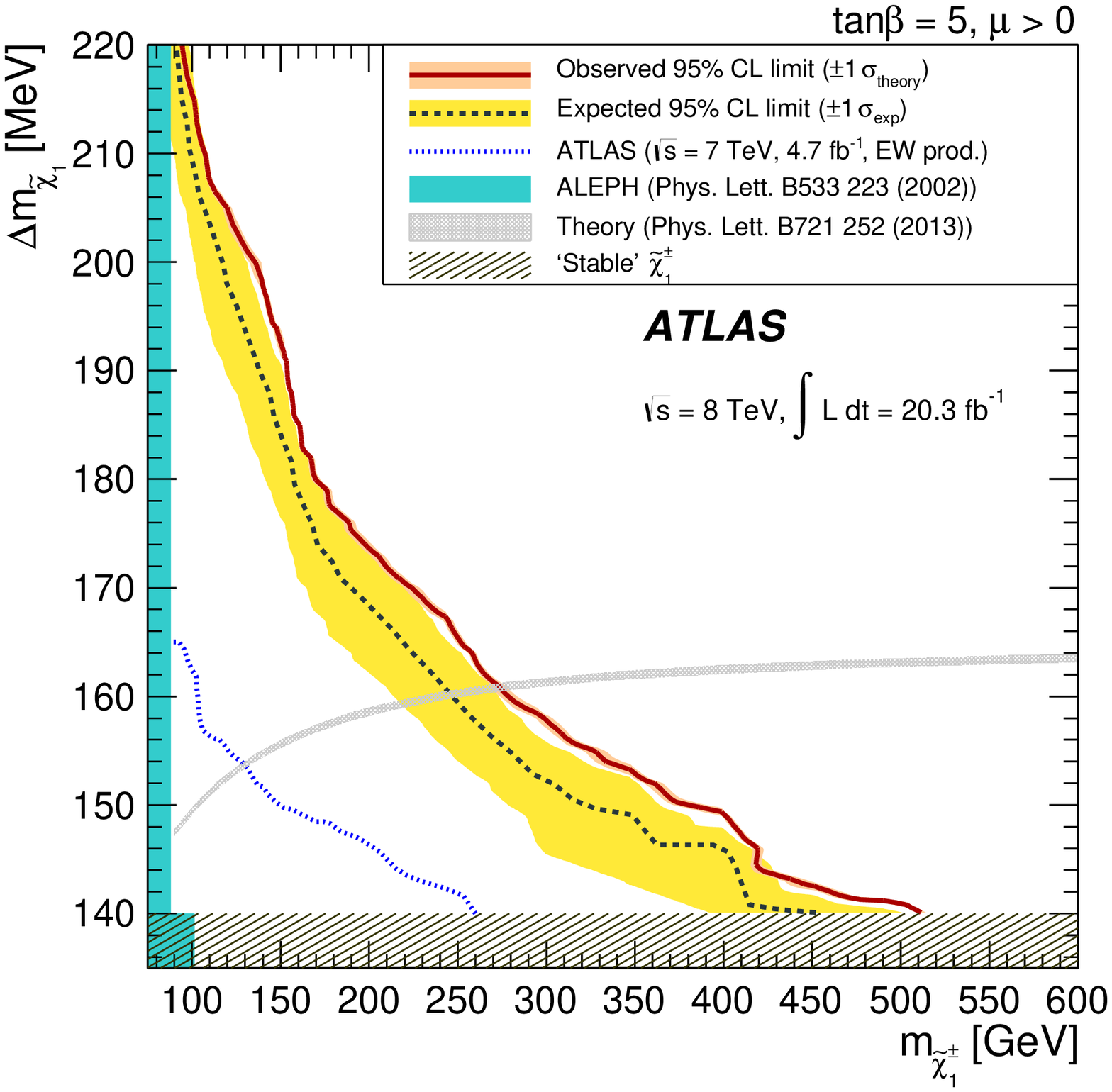}
  \caption{The constraint on the allowed $\mdiff\text{--}m_{\chinoonepm}$ space of the AMSB model for $\tan\beta = 5$ and $\mu>0$.
  The dashed line shows the expected limits at 95\% CL, with the surrounding shaded band indicating the 1$\sigma$ exclusions due to experimental uncertainties.
  Observed limits are indicated by the solid bold contour representing the nominal limit and the narrow surrounding shaded band is obtained by varying the cross-section by the theoretical scale and PDF uncertainties.
  The previous result from Ref.~\cite{ATLAS:2012jp} and an example of the limits achieved at LEP2 by the ALEPH experiment~\cite{Heister:2002mn} are also shown on the left by the dotted line and the shaded region, respectively.
  Charginos in the lower shaded region could have significantly longer lifetime values for which this analysis has no sensitivity as the chargino does not decay within the tracking volume. For this region of long-lived charginos, the limits achieved at LEP2 by the ALEPH experiment is $101~\GeV$~\cite{Heister:2002mn}.}
  \label{figure:limit_dM_mC1}
\end{figure}

\section{Conclusions}\label{sec:conclusion}
The results from a search for charginos nearly mass-degenerate with the lightest neutralino based on the high-$\pt$ disappearing-track signature are presented.
The analysis is based on 20.3 \ifb of $pp$ collisions at $\rts = 8~\TeV$ collected by the ATLAS experiment at the LHC.
The $\pt$ spectrum of observed candidate tracks is found to be consistent with the expectation from SM background processes, and no indication of decaying charginos is observed.
Constraints on the chargino mass, the mean lifetime and the mass splitting are set, which are valid for most scenarios in which the lightest supersymmetric particle is a nearly pure neutral wino.
In the AMSB models, a chargino having a mass below $270~\GeV$ is excluded at 95\% CL.

\begin{acknowledgments}
We thank CERN for the very successful operation of the LHC, as well as the
support staff from our institutions without whom ATLAS could not be
operated efficiently.

We acknowledge the support of ANPCyT, Argentina; YerPhI, Armenia; ARC,
Australia; BMWF and FWF, Austria; ANAS, Azerbaijan; SSTC, Belarus; CNPq and FAPESP,
Brazil; NSERC, NRC and CFI, Canada; CERN; CONICYT, Chile; CAS, MOST and NSFC,
China; COLCIENCIAS, Colombia; MSMT CR, MPO CR and VSC CR, Czech Republic;
DNRF, DNSRC and Lundbeck Foundation, Denmark; EPLANET, ERC and NSRF, European Union;
IN2P3-CNRS, CEA-DSM/IRFU, France; GNSF, Georgia; BMBF, DFG, HGF, MPG and AvH
Foundation, Germany; GSRT and NSRF, Greece; ISF, MINERVA, GIF, DIP and Benoziyo Center,
Israel; INFN, Italy; MEXT and JSPS, Japan; CNRST, Morocco; FOM and NWO,
Netherlands; BRF and RCN, Norway; MNiSW, Poland; GRICES and FCT, Portugal; MNE/IFA, Romania; MES of Russia and ROSATOM, Russian Federation; JINR; MSTD,
Serbia; MSSR, Slovakia; ARRS and MIZ\v{S}, Slovenia; DST/NRF, South Africa;
MINECO, Spain; SRC and Wallenberg Foundation, Sweden; SER, SNSF and Cantons of
Bern and Geneva, Switzerland; NSC, Taiwan; TAEK, Turkey; STFC, the Royal
Society and Leverhulme Trust, United Kingdom; DOE and NSF, United States of
America.

The crucial computing support from all WLCG partners is acknowledged
gratefully, in particular from CERN and the ATLAS Tier-1 facilities at
TRIUMF (Canada), NDGF (Denmark, Norway, Sweden), CC-IN2P3 (France),
KIT/GridKA (Germany), INFN-CNAF (Italy), NL-T1 (Netherlands), PIC (Spain),
ASGC (Taiwan), RAL (UK) and BNL (USA) and in the Tier-2 facilities
worldwide.

\end{acknowledgments}

\providecommand{\href}[2]{#2}\begingroup\raggedright\endgroup

\clearpage 
% ATLAS Collaboration author list
% Data extracted on 23-Aug-2013 for paper reference SUSY-2013-01
%\documentclass[11pt]{article}
%\usepackage{a4wide}\begin{document}
\begin{flushleft}
{\Large The ATLAS Collaboration}

\bigskip

G.~Aad$^{\rm 48}$,
T.~Abajyan$^{\rm 21}$,
B.~Abbott$^{\rm 112}$,
J.~Abdallah$^{\rm 12}$,
S.~Abdel~Khalek$^{\rm 116}$,
O.~Abdinov$^{\rm 11}$,
R.~Aben$^{\rm 106}$,
B.~Abi$^{\rm 113}$,
M.~Abolins$^{\rm 89}$,
O.S.~AbouZeid$^{\rm 159}$
% ...
\bigskip
\\
% ...
$^{11}$ Institute of Physics, Azerbaijan Academy of Sciences, Baku, Azerbaijan\\
$^{12}$ Institut de F{\'\i}sica d'Altes Energies and Departament de F{\'\i}sica de la Universitat Aut{\`o}noma de Barcelona, Barcelona, Spain\\
% ...
$^{21}$ Physikalisches Institut, University of Bonn, Bonn, Germany\\
% ...
$^{48}$ Fakult{\"a}t f{\"u}r Mathematik und Physik, Albert-Ludwigs-Universit{\"a}t, Freiburg, Germany\\
% ...
$^{89}$ Department of Physics and Astronomy, Michigan State University, East Lansing MI, United States of America\\
% ...
$^{106}$ Nikhef National Institute for Subatomic Physics and University of Amsterdam, Amsterdam, Netherlands\\
% ...
$^{112}$ Homer L. Dodge Department of Physics and Astronomy, University of Oklahoma, Norman OK, United States of America\\
$^{113}$ Department of Physics, Oklahoma State University, Stillwater OK, United States of America\\
% ...
$^{116}$ LAL, Universit{\'e} Paris-Sud and CNRS/IN2P3, Orsay, France\\
% ...
$^{159}$ Department of Physics, University of Toronto, Toronto ON, Canada\\
% ...
\end{flushleft}

%\end{document}
% Created with ./xml2latex.py

\end{document}